\documentclass[aip]{revtex4-1}

\usepackage{graphicx}
\draft 
\usepackage{xcolor}

\begin{document}

\title{Extracting Atomic Environments for Machine Learning Interatomic Potentials} 

\author{Jared C. Stimac}
\email[]{stimac1@llnl.gov}

\affiliation{Lawrence Livermore National Laboratory \\
  Livermore, California 94550, USA \\}

\author{Fei Zhou}
\affiliation{Lawrence Livermore National Laboratory \\
  Livermore, California 94550, USA \\}

\author{Kyle Bushick}
\affiliation{Lawrence Livermore National Laboratory \\
  Livermore, California 94550, USA \\}

\author{Bo Lei}
\affiliation{Lawrence Livermore National Laboratory \\
  Livermore, California 94550, USA \\}
  
\author{Sebastien Hamel}
\affiliation{Lawrence Livermore National Laboratory \\
  Livermore, California 94550, USA \\}

\author{Amit Samanta}
\affiliation{Lawrence Livermore National Laboratory \\
  Livermore, California 94550, USA \\}

\author{Vincenzo Lordi}
\affiliation{Lawrence Livermore National Laboratory \\
  Livermore, California 94550, USA \\}

\begin{abstract}
In order to appropriately capture large-scale material features and emergent phenomena via atomistic simulations, such as Molecular Dynamics (MD), the system scale can range up to hundreds of millions of atoms. 
However, the force-field models that drive those simulations are generally trained with Density Functional Theory (DFT) reference data, limited to relatively small configurations on the order of 100s or 1000s of atoms. 
To compute DFT forces on atoms in regions of interest, for example for active-learning or on-the-fly training of interatomic potentials, one needs to extract a small set of atoms from the larger simulation box, and typically work with periodic boundary conditions for DFT. 
However, methods to select the shape and size of this extracted set of atoms, as well as to generate a potentially necessary passivating envelope, have not been systematically analyzed.
In this work, we benchmark several techniques, including a generative diffusion-based artificial intelligence (AI) approach, for extracting atomic environments from large, bulk configurations and embedding them into smaller configurations suitable for DFT calculations with periodic boundary conditions. 
We test with a diverse set of material systems, which includes amorphous $\mathrm{SiO_2}$, Ta with screw dislocations, and molten C. 
We demonstrated a notably simple procedure, a method we refer to as \textit{deletions}, yields superior performance over an array of alternative extraction methods.
\end{abstract}

\pacs{}

\maketitle 

\section{Introduction}
\label{sec:intro}

Developing machine-learned interatomic potentials (MLIPs) poses several challenges, with a prominent one being training data curation.
Ideally, a training dataset is composed of atomic structures that aptly spans the space of configurations that will be sampled during production simulations.
This prevents extrapolation and nonphysical results. 
However, assembling a dataset approaching this ideal is nontrivial, and developing strategies to aid this process constitutes a major area of contemporary research in atomistic modeling.
These strategies generally involve sampling training structures from simulations carried out at relevant conditions; however, additional difficulty arises due to the scaling of calculations used to provide the ground-truth labels (i.e., energies and forces) of the training structures.
Conventional electronic structure methods, the most prominent being Density Functional Theory (DFT), are generally limited to $\mathcal{O}(10^3)$ atoms---even with state-of-the-art resources.
This limitation can be problematic when critical atomic structures only emerge during large-scale simulations with orders of magnitude more atoms than what can be reasonably handled with DFT.
Unless this limitation is resolved, active-learning workflows will fail to incorporate important regions of the energy landscape into the resulting interatomic potential.

One potential solution for the scaling limitation is to extract clusters of atoms within regions of interest from the structures too large for DFT, and reconstruct them within smaller configurations that are DFT tenable. 
This strategy relies on sufficient \textit{locality} of the atomic interactions, the same principle by which interatomic potentials are able to decompose the quantum-mechanical energy into atomic contributions.
This is achieved by constructing atomic energy functions over each atom's local environment and is imperative for scalability.  
Locality is attributed to both charge screening and nearsightedness. 
The latter was rigorously formalized in seminal work by Kohn and Prodan\cite{kohn1996density, prodan2005nearsightedness, prodan2006nearsightedness}, but put simply: the electronic charge density at a point is unaffected by perturbations done far enough away---and in the absence of long-range electrostatics. 
Of course, how far away the perturbation must be and how fast its effect decays spatially depends on both the material system as well as the nature of the perturbation.  

While sufficient locality in a material system is a general requirement for successful atomic-environment extraction, there are many choices that need to be made regarding the implementation, including how far from the atoms of interest should the structure be preserved and whether, and how, defects introduced at the extraction boundary should be removed.
For bulk materials, an added difficulty comes with dealing with periodic boundary conditions (PBC). 
How best to embed or reconstruct an extracted atomic environment in a smaller simulation cell, while accounting for  PBCs, is a nontrivial open problem.  
A few methods have been reported for extracting atomic environments in the context of aiding MLIP data curation. 
Hodapp and Shapeev presented a sophisticated symmetrization procedure, which enabled training a tungsten MLIP with highly accurate dislocations properties \cite{hodapp2020operando}. 
However, this approach has limited generalizability due to the procedure requiring defined ideal lattice sites, precluding its use for amorphous and other less-ordered materials and systems.
Zheng et al., extracted spherical clusters surrounding atoms of interest and embedded those clusters into vacuum, which they used to train a platinum MLIP \cite{zheng2023towards}.
Erhard et al.\ formulated an environment extraction approach referred to as \textit{amorphous matrix embedding} (AME) to aid the construction of a Si-O MLIP. 
In the AME method, new periodic boundaries centered on an atom of interest are imposed and atoms up to a predefined radius are held fixed; 
the configuration is then annealed and quenched to remove spurious environments at the periodic boundaries, requiring potentially time-consuming MD simulations for every extracted environment \cite{erhard2024modelling}.
Goff et al.\ recently introduced the Generalized Representative Structure (GRS) method,\cite{goff2024generalized} in which a large target configuration is represented by smaller configurations generated by converging a distribution of Atomic Cluster Expansion\cite{drautz2019atomic} (ACE) descriptors to that of the target's. 
This last approach is remarkably general, applicable to both crystalline and amorphous systems.

While many previously reported approaches are promising, they are relatively new, and there has been limited work assessing their performance or providing comparative benchmarks, particularly assessing both their accuracy and computational cost.
Moreover, there are no established strategies for their assessment or comparison.
Assessing the quality of an extraction method based on the total energies is problematic; it is of course an extensive property and any decompositions into local or atomic contributions is not uniquely defined. 
By contrast, the atomic forces provide a more direct observation of the local interactions, and, ideally, they should be the same in the extracted clusters as in the larger configuration.  

In this work, we benchmark six atom-centered extraction approaches for bulk materials, half of which are completely novel, including a generative diffusion artificial intelligence (AI) based procedure, 
and also compare to the previously reported AME method (referred to in this work as \textit{anneal}, to be consistent with our nomenclature scheme).
We extract environments from relatively large configurations ($>1,000$ atoms), but which are still amenable to DFT validation, albeit demanding computationally. Thus, we are able to directly compare the DFT-calculated forces before and after extraction as an objective metric.
The goal then is to perfectly reproduce the DFT-computed atomic forces of the extracted atoms within the smaller cells into which they are embedded.
To curb confusion, we note that references in this work to ``accuracies'' or ``errors'' of atomic forces refer to that defined goal.
In addition to evaluating the atomic forces, we also examine pertinent structural aspects and per-atom-averaged energies of the configurations made using the various extractions methods.
We selected three material systems for this benchmark -- amorphous $\mathrm{SiO_2}$, body-centered cubic Ta with screw dislocations, and molten C -- selected to cover a broad range of composition and configuration space of materials.

\section{Methods}

\subsection{Extraction methods}

\begin{figure}[!htb]%
\centering
\fbox{\includegraphics[width=0.95\textwidth]{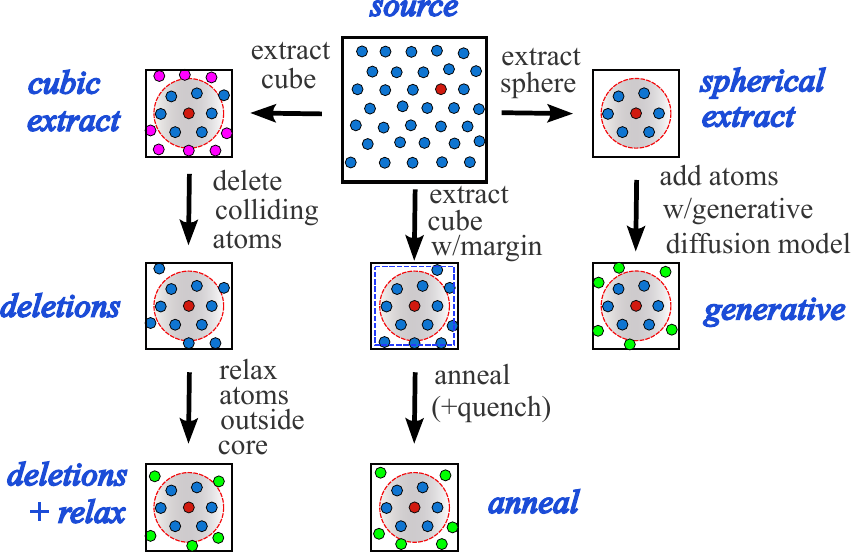}}
\caption{Flow diagram illustrating the six methods for atom-centered environment extractions considered in this work. The method labels, shown with blue, italic text, are adopted throughout.
The fixed atom-centered environment is marked by a central red atom surrounded by a gray area with a red dashed boundary. Blue atoms have equivalent relative positions to the central atom of the extracted cell as in the initial, \textit{source} configuration. Magenta atoms also have equivalent relative position, but are colliding with one another across the periodic boundaries of the \textit{cubic extract} cells. Atoms colored green are either moved from their relative positions or added via a generative diffusion model. In the first step of the \textit{anneal} method, the dotted, blue square denotes the inner edge of a margin in which \textit{source} atoms were not added to discourage collisions across the periodic boundaries.}
\label{fig:flow_diagram}
\end{figure}

Six methods for extracting atom-centered environments are tested in this work, some of which are intermediate steps of others.
In discussing these methods, atoms at the center of the environments selected for extraction are referred to simply as ``central atoms'' (red colored atoms in Fig. \ref{fig:flow_diagram}).
The central atom and those surrounding it, within a predefined fixed sphere, are referred to as the ``fixed core.''
All extraction methods are applied to the same set of fixed cores, 25 per source configuration. 
Additionally, all approaches embed the cores into smaller cubic cells, which we refer to as  ``destination cells''. For a given material system, all destination cells have equal side lengths ($L_{\mathrm{cell}}$), and all fixed cores have the same radii ($r_{\mathrm{core}}$); this was done to enable fair comparisons across extraction methods.
A schematic overview of the extraction methods is shown in Fig. \ref{fig:flow_diagram}, with the parameters for these methods included in Table \ref{tab:method_params}. Detailed descriptions of the methods follows.

\textit{Spherical extract}: In this method, a fixed core, obtained from a source configuration, is placed into the center of an otherwise-empty destination cell. 
The fixed-core cluster is therefore surrounded by a vacuum layer. 
Given the selected destination cell sizes and fixed core radii, the minimum distance between the edge of a fixed core and the edge of a core in a neighboring periodic image is $4.4 \ \mathrm{\AA}$ for $\mathrm{SiO_2}$,  $2 \ \mathrm{\AA}$ for Ta, and  $2 \ \mathrm{\AA}$ for C.

\textit{Generative}: This method first applies the \textit{spherical extract} approach, then
new atoms are added into the vacuum region by a generative diffusion model. 
The new atoms placed into the vacuum region form a coherent structure across periodic boundaries. 
When following this procedure for $\mathrm{SiO_2}$ and C, the generative model would sometimes add new atoms into the fixed core region, in which case those atoms were deleted to maintain the constant fixed core.
This did not occur in the extracted Ta configurations, potentially due to the more regular packing of the metal lattice compared to the liquid C and amorphous $\mathrm{SiO_2}$ structures.
Additional details regarding the diffusion model are included in Sec. \ref{sec:diffusion}. 

\textit{Cubic extract}: In this approach, the fixed core and all of the atoms beyond the fixed core but within the cubic volume of the destination cell are extracted from the source configuration intact.
All atoms in the extracted image have equivalent relative positions with respect to the central atom as they did in the source configuration. 
This procedure is equivalent to imposing new, arbitrary periodic boundaries around a selected atom, which implies that no effort is made to enforce reasonable interactions across the periodic boundaries (\textit{e.g.,} close contacts may occur and are not eliminated). 
 
\textit{Deletions}: This approach starts from configurations generated through the \textit{Cubic extract} method, but then refines the structure by identifying atoms colliding across periodic boundaries by checking the absolute value of the forces on atoms outside the fixed core, to be above some threshold $F_{\mathrm{tol}}$, using a low-computational-cost interatomic potential (IAP) model, preselected for each material. 
The atom with the largest absolute force is removed and the atomic forces are then re-evaluated. 
The process of removing the high-force atoms is continued until the maximum force is less than a predefined tolerance ($\mathrm{max} \  \{F_i\}<F_{\mathrm{tol}}$). 
Values of $F_\mathrm{tol}$ were chosen to be close to, but greater than, the maximum DFT force in the respective source configurations.

\textit{Deletions + relax}: This method starts with a configuration constructed by the \textit{deletions} method.
Atoms outside the core are then relaxed via energy minimization with an IAP. (We chose the BFGS\cite{broyden1970convergence} minimization algorithm, as implemented in the ASE code\cite{ase-paper}.) 
A maximum-force stopping criterion of $0.15 \ \mathrm{eV/\AA}$ was used throughout this work. 
Although this stopping criterion would be considered high for most geometry-optimization applications, the goal here is to remove spurious interactions while producing structures that resemble the source configurations. 
To prevent atoms from entering the fixed core during the energy minimization, a repulsive wall potential was added to the forces from the IAP. 
Details of the repulsive wall are included in Sec. I of the Supporting Information (SI).

\textit{Anneal}: This method closely follows the AME method outlined in Ref. \citenum{erhard2024modelling}. 
First, new periodic boundaries are imposed on a selected central atom, similarly to \textit{cubic extract}, except a small margin ($\delta_{margin} = 0.5~\mathrm{\AA}$) is enforced between the region of extracted atoms and the boundary of the destination cell.
This margin helps mitigate collisions across periodic boundaries.
In the case of SiO$_2$, we remove as few atoms as needed from the region outside of the fixed core in order to achieve the desired stoichiometry, just as was done in Ref. \citenum{erhard2024modelling}. 
Next, atoms outside the fixed core are annealed using an IAP at 4000~K for 10 or 50~ps, depending on the material, as specified in Table \ref{tab:method_params}. 
For SiO$_2$ and Ta, structures were then quenched to 300~K with quench rate of $3.7 \times 10^{12}$~K/s. For C, the structures were not quenched, to produce configurations more consistent with the source configuration, which was sampled from a high-temperature MD trajectory.

\begin{table}[h]
\caption{Parameters of the extraction methods that varied across materials systems, which were not specified elsewhere. This includes the side lengths of the cubic destination cells that the extracted environments are placed into  ($L_{\mathrm{cell}}$); radii used to define the fixed core regions ($r_{\mathrm{core}} $); tolerance for absolute value of forces used for the \textit{deletions} method; time the structure was annealed during the \textit{anneal} method ($t_{\mathrm{anneal}}$).}
\label{tab:method_params}
\begin{tabular}{lccc}
\toprule%
                                       &       $\mathrm{SiO_2}$      &     Ta      &        C      \\ 
\hline
$L_{\mathrm{cell}} \ [\mathrm{\AA}]$    &           14.4              &      14.0   &       14.0    \\
 $r_{\mathrm{core}} \ [\mathrm{\AA}]$   &           5.0                &    6.0     &         6.0   \\
$F_{\mathrm{tol}} \ [\mathrm{eV/\AA}] $ &           7.5                &    10     &         25   \\
$t_{\mathrm{anneal}} \ [\mathrm{ps}]$   &          50                &      50      &       10 \\
\end{tabular}
\end{table}

\subsection{Generative diffusion model}
\label{sec:diffusion}
Generative AI has recently emerged as a powerful paradigm for atomic structure generation in computational materials science \cite{xie2021crystal, zheng2023towards, Kwon2023MLST}, offering a way to sample high-dimensional configuration spaces that are difficult to explore with traditional simulation methods. Diffusion models \cite{Sohl-Dickstein2015-DPM, Ho2020-DDPM, Song2020-unified} , in particular, provide a principled framework for learning the underlying probability distribution of atomic arrangements and generating new structures through iterative denoising guided by a learned score field. We adopted the method developed by Kwon et al. \cite{Kwon2023MLST}, which demonstrated accurate and scalable generation of relatively large supercells. 
Following this framework, we trained a diffusion-based score model on a curated dataset of atomic structures spanning the relevant structural motifs.  
The diffusion models for $\mathrm{SiO_2}$ and C were trained using the entire training sets that were used to make the IAPs in Refs. \citenum{erhard2024modelling} and \citenum{lindsey2017chimes}, respectively. 
For the Ta model, six frames were sampled from the compression simulation used to generate the source configurations; they were sampled uniformly between the initially unstrained starting configuration to a true strain of $-1.0\%$. 

For inference, we adopted an outpainting protocol: beginning from a fixed core cluster of atoms, we randomly initialized the positions of atoms to be generated in the surrounding region and evolved them through the reverse-diffusion process using the trained score model. This guided denoising gradually reconstructs physically plausible atomic configurations consistent with the learned structural distribution, enabling controlled expansion of the initial cluster into a full periodic structure. For each given core structure, we varied the number of atoms to generate and picked the top ten most ``reasonable'' structures by comparing the sampled and ground truth radial distribution functions.

\subsection{Source configurations}
\begin{figure}[!htb]%
\centering
\includegraphics[width=0.90\textwidth]{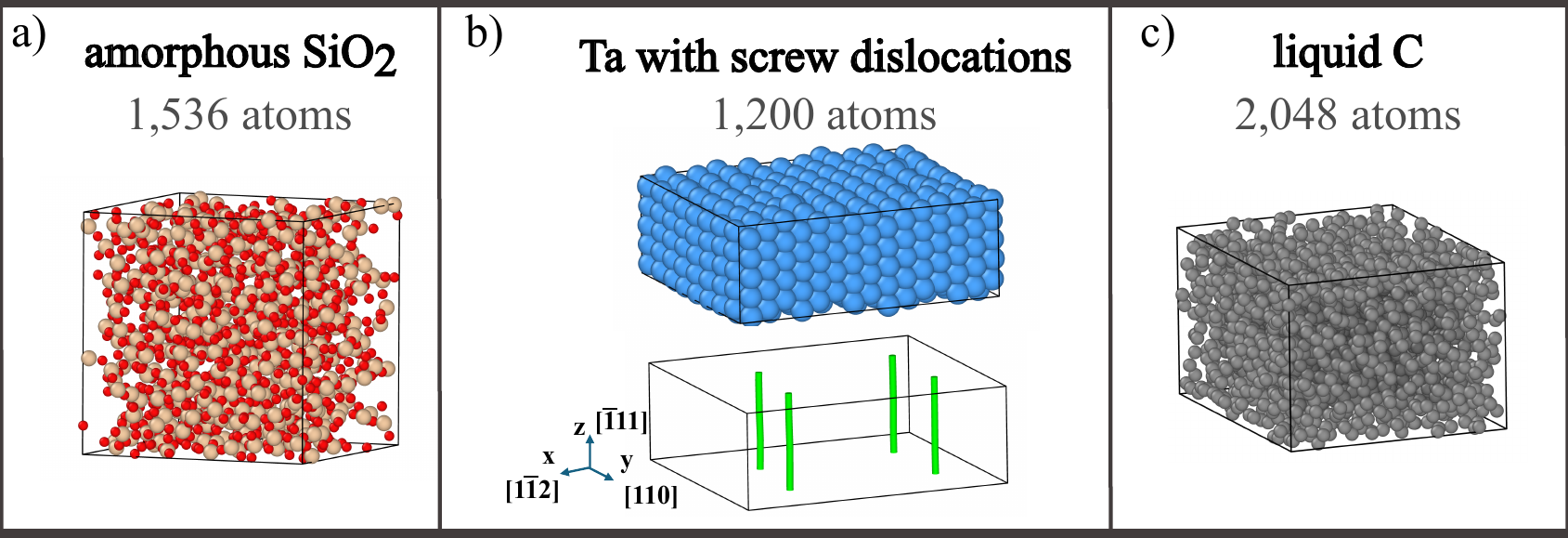}
\caption{Source configurations for (a) $\mathrm{SiO_2}$, (b) Ta, and (c) C.}
\label{fig:source_configs}
\end{figure}

Three different configurations, one per material, were constructed as sources from which to extract atomic environments (Fig. \ref{fig:source_configs}).
As these are all bulk structures, they were generated with simulations using full PBCs. 
The first configuration is amorphous $\mathrm{SiO_2}$  with 1536 atoms at a density of $2.2 \ \mathrm{g/cm^3}$, sampled from an MD simulation with fixed particle number, volume, and temperature (NVT) at 300~K (Langevin thermostat).
The second source configuration comprises body-centered cubic (BCC) Ta, with 1200 atoms and a quadrupole of screw dislocations, generated using Atomsk\cite{hirel2015atomsk}. 
The dislocation lines are all parallel with the $[\bar{1}11]$ direction, which is aligned with the $z$ axis. 
The signs of the Burgers vectors are alternating so that the dislocations do not break the periodicity of the lattice along the $x$ and $y$ axes. 
The configuration was taken from a frame of a constant strain rate ($-2 \times 10^8 \ \mathrm{s^{-1}}$) compression MD simulation using a Langevin thermostat with the temperature set to 300 K. 
The compression was applied along the $y$ axis, while the $x$ and $z$ axes were expanded to maintain a constant volume throughout the simulation. 
The source configuration was sampled at a point in the simulation with a true strain of $0.6\%$, just prior to stress-induced annihilation of the dislocations.
The Ta atoms selected as `central atoms' for the environment extraction tests, meaning those at the center of the small destination cells, were sampled across the source configuration to capture a range of local environments, including both perfect crystal and dislocation core regions (see Supporting Information Sec.~III).
The last source configuration is molten C with 2048 atoms at $2.43 \ \mathrm{g/cm^3}$, which was sampled from an NVT simulation at 5000~K. 

\subsection{Interatomic potentials (IAPs)}
An IAP was required for each material system in order to construct the source configurations and for various steps in some of the extraction methods.
The IAPs were taken from previously reported studies and selected based on the targeted conditions of the source configurations. 
For $\mathrm{SiO_2}$, an atomic cluster expansion (ACE)\cite{drautz2019atomic} MLIP model was used that was reported in the same publication that introduced the AME method\cite{erhard2024modelling}; this was done to reduce systematic differences between AME as done in that work and the \textit{anneal} method here.
For Ta, we used an embedded atom method (EAM) potential reported by Li et al.\cite{li2003embedded}
For molten C, a Chebyshev Interaction Model for Efficient Simulation (ChIMES) MLIP reported by Lindsey et al. \cite{lindsey2017chimes} was used. 

\subsection{DFT calculations}
All DFT calculations were performed using the Simulation Package for Ab-initio Real-space Calculations (SPARC) software\cite{zhang2024sparc} and the Perdew-Burke-Ernzerhov (PBE) \cite{perdew2008restoring} generalized gradient approximation (GGA) exchange-correlation functional. 
Finite-difference mesh sizes of $0.08$, $0.15$, and $0.10 \ \mathrm{\AA}$ were used for $\mathrm{SiO_2}$, Ta, and C, respectively.
All calculations of the source configurations used a $\Gamma$-point only $k$-point mesh, while those of the destination cells used $2\times2\times2$, for all materials. 
Convergence testing for the finite-difference and $k$-point meshes is included in Appendix~\ref{sec:dft_convergence}.
Fermi-Dirac smearing was used with the effective electronic temperatures set  equal to the corresponding ionic temperatures of the simulations from which the configurations were sampled: $9.5\times 10^{-4}$, $9.5\times 10^{-4}$, and $1.583\times 10^{-2}$~Ha, for $\mathrm{SiO_2}$, Ta, and C, respectively. 
The D3 dispersion correction by Grimme \cite{grimme2010consistent} was used for all calculations with C.

\section{Results}
\subsection{Accuracy of forces for the central atoms}
\begin{figure}[!htb]%
\centering
\fbox{\includegraphics[width=0.90\textwidth]{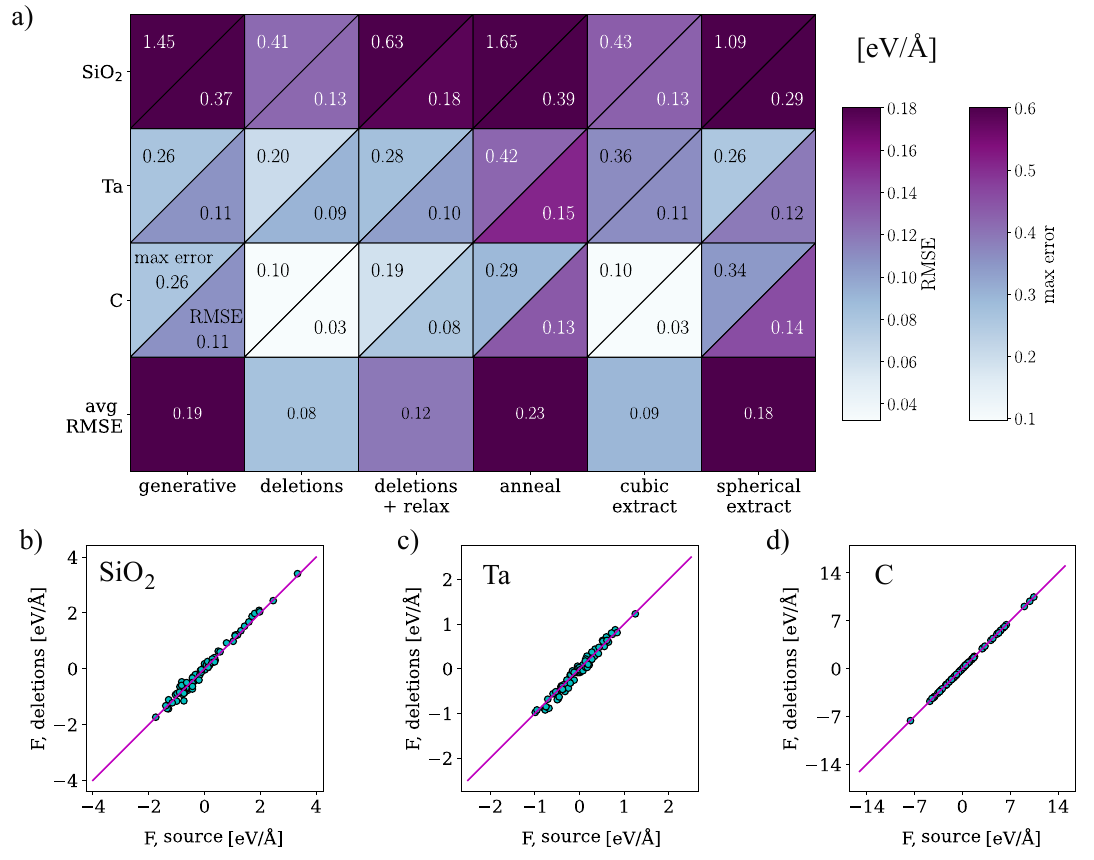}}
\caption{Results for the atoms at the center of the extracted environments. (a) Maximum absolute error and root-mean-squared error (RMSE) of the atomic forces for each material and extraction method. Parity plots of forces for atoms from the source configuration versus those in the \textit{deletions}-extracted configurations for (b) $\mathrm{SiO_2}$, (c) Ta, and (d) C.}
\label{fig:central_atom_force_accuracy}
\end{figure}

We first consider the ability of the extraction methods to reproduce the atomic forces for only the atom at the center of each extracted cell. 
In an extracted cell, the central atom retains the largest local environments equivalent to that in the source configuration, so it should give the best performance in terms of reproducing the atomic forces. 
Figure~\ref{fig:central_atom_force_accuracy} presents the accuracy of the atomic forces evaluated exclusively for the single atoms at the center of the extracted environments.
Again, the DFT-computed forces from the source configurations serve as the reference.
Of the six extraction methods tested here, the \textit{deletions} method produced the lowest average root-mean-squared error (RMSE) at $0.08 \ \mathrm{eV/\AA}$, while the \textit{anneal} method resulted in the highest at $0.23 \ \mathrm{eV/\AA}$.  
Moreover, for any of the three materials, \textit{deletions} was either the best or tied for best for both RMSE and maximum absolute error (max error). 
The \textit{cubic extract} method gave the closest next-best performance to \textit{deletions}, with equivalent RMSE for the $\mathrm{SiO_2}$ and C forces. 

Considering just the RMSEs and max errors alone is insufficient to determine whether these trends would generalize to larger sample populations.
To assess whether differences in the error distributions are statistically significant, or simply reflect limited sample sizes, we performed paired Student t-tests on the squared-residual errors for all pairs of the extraction methods.
The test considers the value of $p$, the probability that the two distributions would occur assuming a Null hypothesis that the values were sampled from equivalent distributions, and we use the standard criteria for rejecting the Null hypothesis:  $p\leq0.05$.
The full analysis is in Appendix \ref{sec:stat_signif}, but we note the most relevant $p$-values here.
First, based on this analysis, we find that the difference in performance between the \textit{deletions} and \textit{cubic extract} is not statistically significant; for example, we find $p=0.80$ for the $\mathrm{SiO_2}$ force errors.  
In contrast, the differences between error distributions of \textit{deletions} and those of all other methods are statistically significant, with a few exceptions.
Specifically, for the Ta environments, the error distributions from \textit{generative} and \textit{deletions + relax} are not significantly different from that of \textit{deletions} with $p=0.08$ and $p=0.16$, respectively. 

Figures~\ref{fig:central_atom_force_accuracy}(b)--(d) show the forces from the \textit{deletions} method versus those from the source, delineated by material.
The relative accuracy has a notably strong dependence on the material system. 
In particular, C shows by far the best relative error.
This can potentially be attributed to the high temperature liquid phase's inherent lack of strong directional bonding, which would make the material less affected by perturbing ionic positions or making arbitrary slices, compared to Ta and $\mathrm{SiO_2}$. 
 We also used a high effective electronic temperature for the C DFT, which is justified by the ionic temperature ($5,000$ K) of the MD simulation from which the source configurations were sampled, but which also contributes to the strong local character for that system.\cite{suryanarayana2017nearsightedness}

\subsection{Accuracy of forces for atoms within the fixed cores}

\begin{figure}[!htb]%
\fbox{\includegraphics[width=0.95\textwidth]{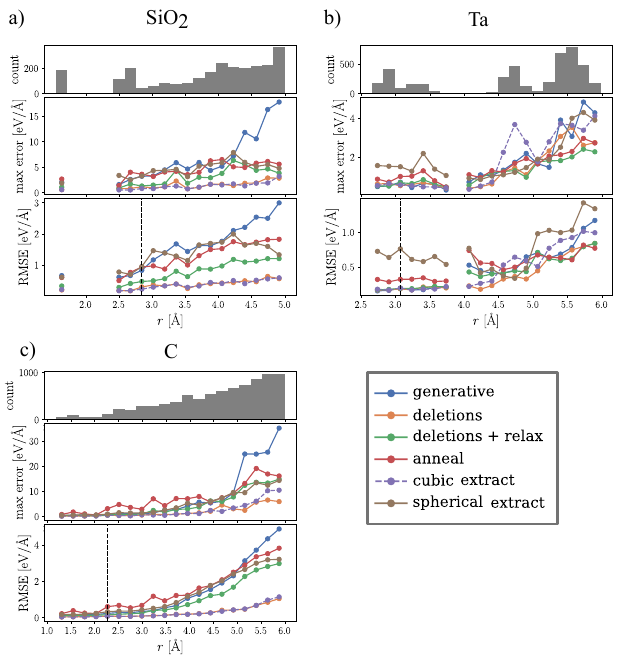}}
\caption{Accuracy of forces for non-central atoms within $r_c$ of the extracted cells for (a) $\mathrm{SiO_2}$, (b) Ta, and (c) C. Atoms are binned based on their distance from the central atom, $r$, and the total count, max error, and root-mean-squared error for all of the atomic forces in each bin are reported. The vertical black dashed lines denotes the radius of the first bin at which the RMSE for \textit{deletions} exceeds twice the RMSE of the central atoms.}
\label{fig:fixed_core_forces}
\end{figure}

So far we have only examined the accuracy of the central atom forces, however there is considerable interest in utilizing more of the DFT-computed forces than just those of the central atom, \textit{e.g.}, when using an extracted environment to augment the training of a MLIP.
If there is a close cluster of atoms in a source configuration for which the forces are needed, ideally, one would only need to extract and perform DFT once, rather than extracting and running DFT on slightly different destination cells individually, improving throughput and data density for active learning.
As one considers atoms farther from the center of a destination cell, the sizes of their local environments that are unchanged from what they experienced in the source configuration gets smaller as they near the newly imposed periodic boundaries. 
This means that atoms too far from the center will have local environments, and therefore forces, poorly reflecting what would be found in the source.
Therefore, we next analyze the accuracy of forces for all other atoms in the fixed core, and how the errors increase as a function of distance from the center.
Figure~\ref{fig:fixed_core_forces} shows these results for each extraction method and material. 

By and large, the same trends found for the central atoms extend to the other atoms in the fixed core; namely, that \textit{deletions} and \textit{cubic extract}  generally show lower force error at a given distance from the cell centers.
This is most notable for $\mathrm{SiO_2}$ and C, for which \textit{deletions} and \textit{cubic extract} demonstrated remarkably gradual error increases away from the center compared to the other methods, resulting in notable gaps between the RMSEs for these methods compared to all other methods, when looking at atoms further away from the fixed cores ($\gtrsim 3.25$~\AA).
However, those extraction methods were less obviously advantageous for Ta at larger distances, while still remaining among the highest performing methods and superior at close distances ($\lesssim 4.5$~\AA). 
In particular, for Ta, for the first and second nearest neighbor (NN) shells ($\sim 2.7 - 3.7\ \mathrm{\AA}$), \textit{deletions + relax} and \textit{generative}  show comparable RMSEs and max errors to \textit{deletions} and \textit{cubic extract}; 
for the third NN shell of Ta ($\sim 4.1 - 5.1\ \mathrm{\AA}$), there are larger fluctuations of the force errors as a function of distance from the central atom, which makes ranking methods less clear (although \textit{deletions} generally has the lowest RMSE), however the point is mostly moot since the magnitudes of the errors are quite large already (RMSEs approaching $\sim 0.5 \ \mathrm{eV/\AA}$ or higher by 4.5~\AA).

\subsection{Structural analysis}
\begin{figure}[!htb]%
\includegraphics[width=0.90\textwidth]{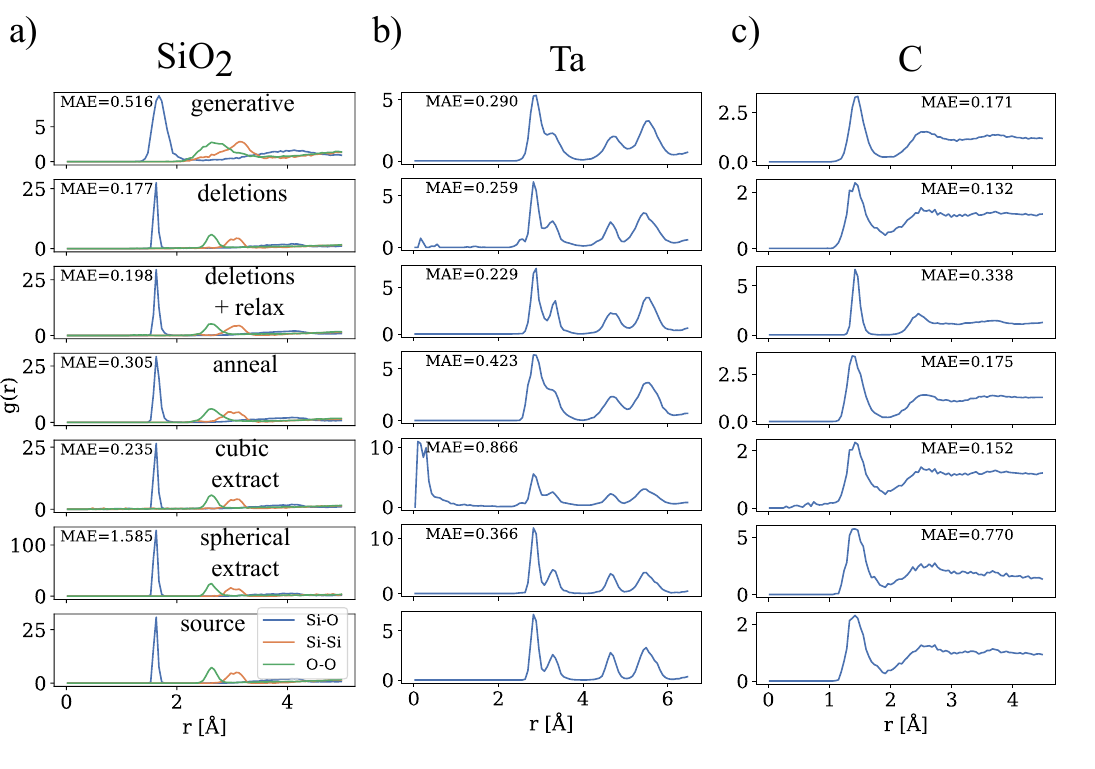}
\caption{Radial distribution functions, $g(r)$, for all extraction methods and the source configurations for (a) $\mathrm{SiO_2}$, (b) Ta, and (c) C. For all extraction methods except \textit{spherical extract}, only atoms outside the fixed cores were used as reference atoms. Each extraction method's subfigure reports the mean-absolute error (MAE) between its $g(r)$ and that of the source configurations.}
\label{fig:rdfs}
\end{figure}

To assess the structural effects that result from the different extraction methods, we examine the radial distribution functions (RDFs).
In Fig. \ref{fig:rdfs}, we report the RDFs averaged over all configurations for a given extraction method, as well as the RDFs for the source configurations as references. For all methods, except \textit{spherical extract}, the RDFs are calculated using only the atoms outside the fixed cores as reference atoms to focus the analysis on regions contributed by the extraction methods. Because \textit{spherical extract} contains no atoms outside the fixed cores, all atoms within the cores are used as references atoms instead.

Inspecting the RDFs for $\mathrm{SiO_2}$, we find that all of the extraction methods give qualitative agreement with the source configuration, except for that of \textit{generative}, which has noticeably broadened peaks. 
We reason this due to our choice to use the entire MLIP training set from Ref. \citenum{erhard2024modelling} to train the underlying diffusion model for that extraction approach. 
Unlike the diffusion models for Ta and C, the $\mathrm{SiO_2}$ model was trained on data containing several other phases and densities beyond what composed its respective source configuration.  
While the diffusion model was inherently more general in this case, this seems to decrease the generated atoms' similarity to the source.
Put differently, the more targeted training data used for Ta and C allowed the generative approach to add atoms to be better aligned with the source configuration, but at the cost of covering narrower regions of configuration space than in the $\mathrm{SiO_2}$ case.
Both targeted and broader training sets are reasonable scenarios to consider here, since different MLIPs span different extents of phase space, as reflected in the underlying IAPs used in this work. 

Additionally, although difficult to see in Fig. \ref{fig:rdfs} (see SI section V for zoomed-in figures), the \textit{cubic extract} RDF contains an extraneous peak near $r=0$, as the procedure does not prevent atoms from colliding across the periodic boundaries. 
Another feature we notice, is the scale of the RDF from \textit{spherical extract} is increased due to the lower density from the large vacuum region around the cell perimeters; this effect occurs in the other materials as well.

\begin{figure}[!htb]%
\includegraphics[width=0.25\textwidth]{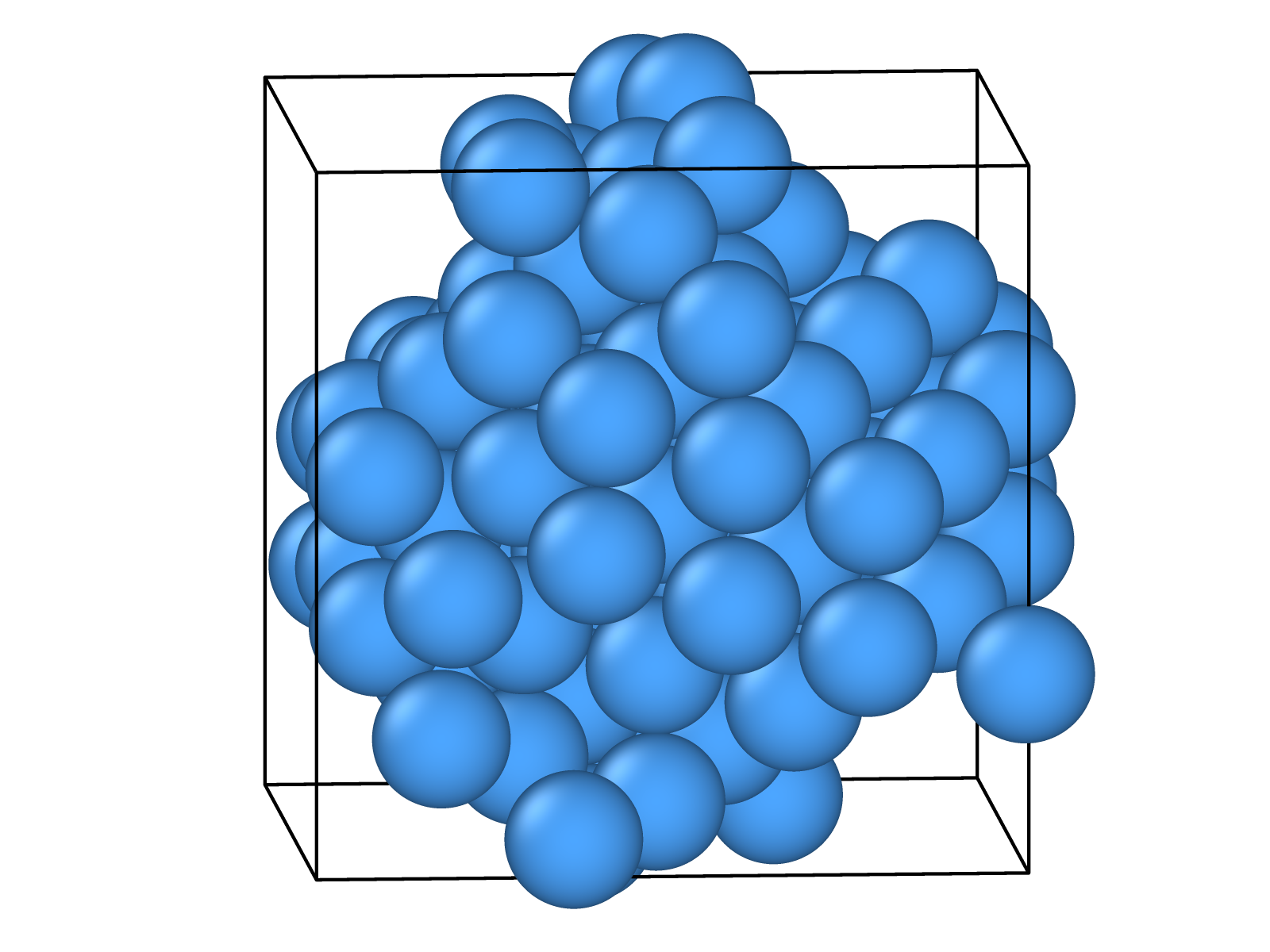}
\caption{Example Ta structure from the \textit{anneal} extraction method.}
\label{fig:Ta_anneal_config}
\end{figure}

For Ta, we likewise find the extraction methods yield reasonable agreement on the RDFs with the source, with a few exceptions. 
Just as was found with $\mathrm{SiO_2}$, \textit{cubic extract} gives extraneous peaks near $r=0$; however, with Ta they are much more prominent due to the smaller quantity of atoms per configuration.
Moreover, the RDF for \textit{deletions} also shows small peaks close to $r=0$, meaning the procedure did not fully remove all atomic collisions in that case.
For \textit{anneal}, the first and second NN peaks are broadened compared to those in the source configuration; to a lesser degree, this broadening also occurs with \textit{generative}. 
Examining the configurations produced by the \textit{anneal} method, we find that the quench, the last step of that extraction method's procedure, left small pores of a few atomic radii; an example of one of these configurations is included in Fig. \ref{fig:Ta_anneal_config}. 
With \textit{anneal}, we followed the approach of Ref. \citenum{erhard2024modelling}, starting with a small empty margin around the extraction cells to prevent issues with close atomic collisions. 
This led to the average density of the \textit{anneal} configurations to be 33\% lower than the source configuration, which likely promoted the pore formation (all densities are reported in Supporting Information Sec. IV).
Modifying the quench rate and margin width to mitigate this effect is likely feasible; however, doing so is probably of limited value given that the simple procedures \textit{deletions} and \textit{cubic extract} yielded better force accuracy than alternative methods, including \textit{generative} and \textit{deletions + relax}, which appear to avoid anomalous boundary interactions.

For the C RDFs, we again observe general qualitative agreement between all extraction methods and source except for the first NN peak from the \textit{deletions + relax} method is tightened compared to that of source. 
Also, as found with $\mathrm{SiO_2}$, the \textit{cubic extract} method leads to small features corresponding to collisions across boundaries (see SI Sec. V).

\begin{figure}[!htb]%

\includegraphics[width=0.40\textwidth]{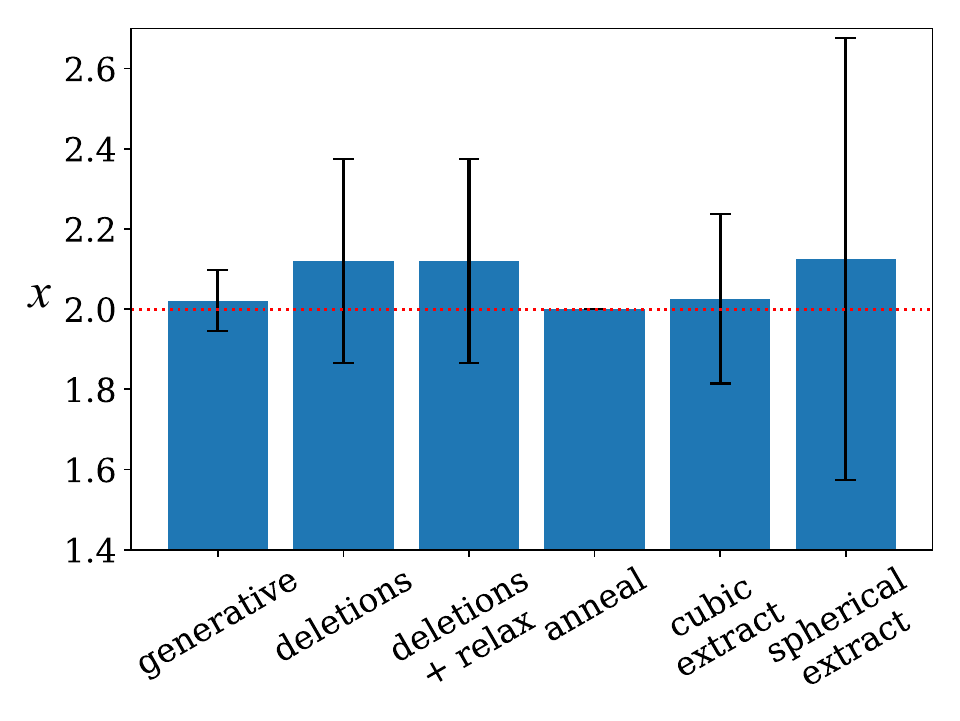}
\caption{Stoichiometric ratio $x$, of the SiO$_x$ extracted configurations. The red horizontal line denotes the stoichiometry of the source configuration. The bar heights are the sample averages while the error bars depict a 95\% confidence interval.}
\label{fig:stoich}
\end{figure}

In Fig. \ref{fig:stoich} we report the stoichiometry of the extracted configurations for the only multi-element system considered here, $\mathrm{SiO_2}$.
The stoichiometry was perfectly preserved for the \textit{anneal} because it was enforced following the procedure outline in Ref. \citenum{erhard2024modelling}.
The other methods produced structures with fairly minor average perturbations in the stoichiometry, but with some giving more substantial variance. 
In particular, the standard deviation of the \textit{spherical extract} configurations is $\sim0.4$ and the coefficient of variation is $\sim15\%$.

\subsection{Energetics}
\begin{figure}[!htb]%
\includegraphics[width=0.40\textwidth]{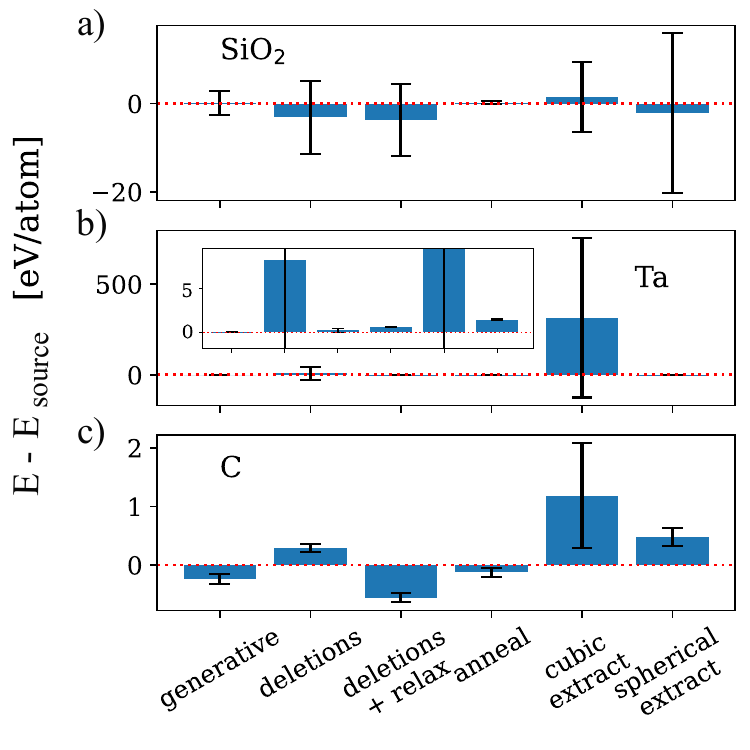}
\caption{Difference between the average energy per atom of the extracted configurations and the source configurations. Error bars reflect 95\% confidence intervals. Inset of (b) is a zoomed in plot of the Ta results with a smaller limit on the y axis.}
\label{fig:energies}
\end{figure}

The DFT total energies of the extracted cells do not make for a rigorous benchmark, as comparing them directly to the total energies of the source configuration is flawed; however, considering the per-atom-averaged energies is of interest as potentially anomalous interactions induced by an extraction method would result in perturbations of this quantity. 
Fig. \ref{fig:energies} shows effects on the energies per atom for all of the extraction methods in reference to their sources. 
The \textit{generative} and \textit{anneal} methods give energies with the most consistent agreement with that of the source configuration. 

Of course, the prevalence of close contacts across PBCs found in the \textit{cubic extract} configurations led to larger per-atom energies than other extraction methods. 
This was most substantial in the case of Ta (Fig. \ref{fig:energies}b), for which we find the energy per atom to be hundreds of electron volts greater than in the source. 
Including such large magnitude energies in the training data of an MLIP is probably not preferable as it could degrade the accuracy of the more physically pertinent regions of configuration space. 
As noted when discussing the RDFs, the \textit{deletions} method did not completely remove all atomic close contacts. 
This also resulted in energies of the Ta configurations using \textit{deletions} to be moderately large ($\sim8$ eV/atom) compared to the source configuration. 
If this did pose an issue when fitting energies, one could make adjustments in how atoms are deleted to amend this. Alternatively, training predominantly on forces is another viable approach, along with use of tailored well-behaved structures to train energies.

\section{Conclusion}
The primary finding in this work is that extraction methods that preserved the central atoms' exact local environments to further radii, close to the boundaries of the cells into which they were embedded, resulted in more accurate atomic forces compared to methods that sought to produce more coherent atomic arrangements across the periodic boundaries.
Various extraction methods produced configurations remarkably consistent structurally with the sources, as illustrated by the RDFs, stoichiometry and per-atom average energies; however, this did not ensure better results on our target of reproducing the forces within the fixed environments.
In particular, \textit{cubic extract} and \textit{deletions} performed the best despite inducing structural inconsistencies at the boundaries.

Although the forces given by the \textit{cubic extract} method were competitive with \textit{deletions}, the excessively large energies undermine its favorability.
For this reason, we assert that the \textit{deletions} approach is, unexpectedly, the optimal extraction method out of those considered---at least for the materials systems examined in this work, although they  represent a reasonably broad range of bulk materials (covalent to metallic bonding with different coordination).
Moreover, the \textit{deletions} method is serendipitously a simple approach to implement and run. 
Unlike other more sophisticated (and computationally costly) approaches we explored, \textit{deletions} did not require constructing additional AI models or running timely MD simulations; the basis of the approach was merely to extract all atoms within a cubic volume, apply PBCs, and delete atoms to remove  collisions across the boundaries of the cells. 
In fact, although we used pretrained MLIPs to identify atomic collisions, one could likely just choose a pair-wise interatomic distance as a tolerance for classifying collisions. 

While we do not include here property prediction benchmarks based on fitting MLIPs to the data generated by the extraction methods, atomic forces are a fundamentally more direct and informative metric of extraction performance than properties predicted by MLIPs trained on data from the different extraction procedures. 
Such comparisons would introduce additional confounding factors, such as functional dependence of the potential and weighting of the training data, along with the size and diversity of the initial training set compared to additionally extracted environments.  
Nonetheless, a more detailed examination of the effects of fitting total energies from the extracted configuration is a potential area of future work. 

Overall, this work suggests a promising path toward improving the construction of MLIP training sets, and therefore, the ability to model large-scale, emergent phenomena in materials.

\section*{Author declarations}

The authors have no conflicts to disclose. 

\section*{Acknowledgments}

This work was performed under the auspices of the U.S. Department of Energy by Lawrence Livermore National Laboratory under Contract DE-AC52-07NA27344, funded by the Laboratory Directed Research and Development Program at LLNL under project tracking code 23-SI-006.

\section*{Supplementary information}
The Supporting Information for this work contains additional details of the repulsive wall used by some of the extraction methods; computational cost of the extraction methods; and magnified view of the $\mathrm{SiO_2}$ and C RDFs.

\section*{Author contributions}
\textbf{JS, FZ}: Methodology, Formal analysis, Software,  Writing - original draft.
\textbf{KB}: Methodology, Interpretation, Writing - review \& editing.
\textbf{BL}: Software, Formal analysis.
\textbf{SH, AS}: Interpretation, Writing - review \& editing.
\textbf{VL}: Funding Acquisition, Project Administration, Supervision, Writing - review \& editing. 

\section*{Data Availability}
The data that supports the findings of this work are included throughout the manuscript and the Supporting Information. Any additional information can be made available upon reasonable request from the corresponding author.

\bibliography{bibliography}

\newpage
\appendix

\renewcommand{\thefigure}{A\arabic{figure}}
\setcounter{figure}{0}
\renewcommand{\thetable}{A\arabic{table}}
\setcounter{table}{0}

\section{Convergence Testing of DFT Calculations}
\label{sec:dft_convergence}

Convergence testing for SiO$_2$ was done with a 3-formula quartz configuration obtained from the training set of Ref. \citenum{erhard2024modelling}; the lattice vectors of the configuration were lengths 4.77, 4.72, 5.29; the SPARC default gaussian smearing with width $0.001 \ \mathrm{Ha}$ was used.

\begin{table}[h!]
\centering
\begin{tabular}{|c|c|c|c|}
\hline
test index, $i$ &  k-mesh & $h$ [Å] & \textbf{$\mathrm{max} \ |F_i - F_{(i-1)}|$} [eV/Å] \\
\hline
0 & $12 \times 12 \times 12 $ & 0.1 & -\\
1 & $12 \times 12 \times 12 $ & 0.09 & 0.019 \\
2 & $12 \times 12 \times 12 $ & 0.08 & 0.002\\
\hline
\end{tabular}
\caption{Convergence tests of the real-space mesh size $h$ for SiO$_2$. }
\label{tab:C_h_mesh}
\end{table}

\begin{table}[h!]
\centering
\begin{tabular}{|c|c|c|c|}
\hline
test index, $i$ &  k-mesh & $h$ [Å] & \textbf{$\mathrm{max} \ |F_i - F_{(i-1)}|$} [eV/Å] \\
\hline
0 & $4 \times 4 \times 4 $ & 0.08 & -\\
1 & $6 \times 6 \times 6 $ & 0.08 & 0.0012\\
2 & $8 \times 8 \times 8 $ & 0.08 & 0.0007\\
\hline
\end{tabular}
\caption{Convergence tests of the k-mesh size $h$ for SiO$_2$. }
\label{tab:C_h_mesh}
\end{table}
For Ta, we tested the DFT parameters using a 2 atom cubic cell with box lengths 3.3 Å, Fermi-Dirac smearing at 300K. The ions were randomly (uniform distribution with magnitude 0.01 Å) perturbed from the energy minimum positions.
\begin{table}[h!]
\centering
\begin{tabular}{|c|c|c|c|}
\hline
test index, $i$ &  k-mesh & $h$ [Å] & \textbf{$\mathrm{max} \ |F_i - F_{(i-1)}|$} [eV/Å] \\
\hline
0 & $10 \times 10 \times 10 $ & 0.20 & -\\
1 & $10 \times 10 \times 10 $ & 0.15 & 0.005\\
2 & $10 \times 10 \times 10 $ & 0.1 & 0.005\\
\hline
\end{tabular}
\caption{Convergence tests of the real-space mesh size $h$ for Ta. }
\label{tab:Ta_h_mesh}
\end{table}

\begin{table}[h!]
\centering
\begin{tabular}{|c|c|c|c|}
\hline
test index, $i$ &  k-mesh & $h$ [Å] & \textbf{$\mathrm{max} \ |F_i - F_{(i-1)}|$} [eV/Å] \\
\hline
0 & $6 \times 6 \times 6 $ & 0.15 & -\\
1 & $8 \times 8 \times 8 $ & 0.15 & 0.008\\
2 & $10 \times 10 \times 10 $ & 0.15 & 0.001\\
\hline
\end{tabular}
\caption{Convergence tests of the k-points mesh size $h$ for Ta. }
\label{tab:Ta_k_mesh}
\end{table}

For C, convergence testing done using a configuration generated with the \textit{deletions} extraction method; the extracted cell was cubic with a side length of 7 Å, containing 43 atoms. Again, default Gaussian smearing (width $0.001 \ \mathrm{Ha}$) was used.

\begin{table}[h!]
\centering
\begin{tabular}{|c|c|c|c|}
\hline
test index, $i$ &  k-mesh & $h$ [Å] & \textbf{$\mathrm{max} \ |F_i - F_{(i-1)}|$} [eV/Å] \\
\hline
0 & $4 \times 4 \times 4 $ & 0.15 & -\\
1 & $4 \times 4 \times 4 $ & 0.1 & 0.060\\
2 & $4 \times 4 \times 4 $ & 0.08 & 0.009\\
\hline
\end{tabular}
\caption{Convergence tests of the real-space mesh size $h$ for C. }
\label{tab:C_h_mesh}
\end{table}

\begin{table}[h!]
\centering
\begin{tabular}{|c|c|c|c|}
\hline
test index, $i$ &  k-mesh & $h$ [Å] & \textbf{$\mathrm{max} \ |F_i - F_{(i-1)}|$} [eV/Å] \\
\hline
0 & $2 \times 2 \times 2 $ & 0.1 & -\\
1 & $4 \times 4 \times 4 $ & 0.1 & 0.205\\
2 & $6 \times 6 \times 6 $ & 0.1 & 0.0150\\
\hline
\end{tabular}
\caption{Convergence tests of the k-points mesh size $h$ for C.}
\label{tab:C_h_mesh}
\end{table}

\newpage
\renewcommand{\thefigure}{B\arabic{figure}}
\setcounter{figure}{0}
\renewcommand{\thetable}{B\arabic{table}}
\setcounter{table}{0} 
\section{DFT vs IAP agreement}

\begin{figure}[!htb]%
\centering
\fbox{\includegraphics[width=0.95\textwidth]{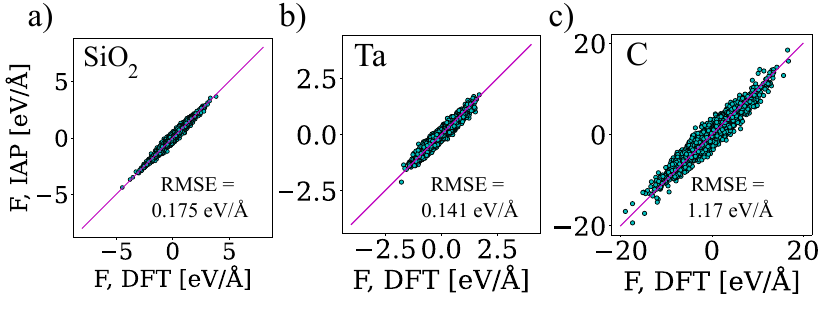}}
\label{}
\caption{Agreement between forces for all atoms within the \textit{extended} configurations given by DFT and the IAPs used for constructing \textit{extended} configurations and in some of the extraction methods.}
\end{figure}

\renewcommand{\thefigure}{C\arabic{figure}}
\setcounter{figure}{0}
\renewcommand{\thetable}{C\arabic{table}}
\setcounter{table}{0} 

\section{Statistical significance}
\label{sec:stat_signif}
\begin{figure}[!htb]%
\fbox{\includegraphics[width=0.95\textwidth]{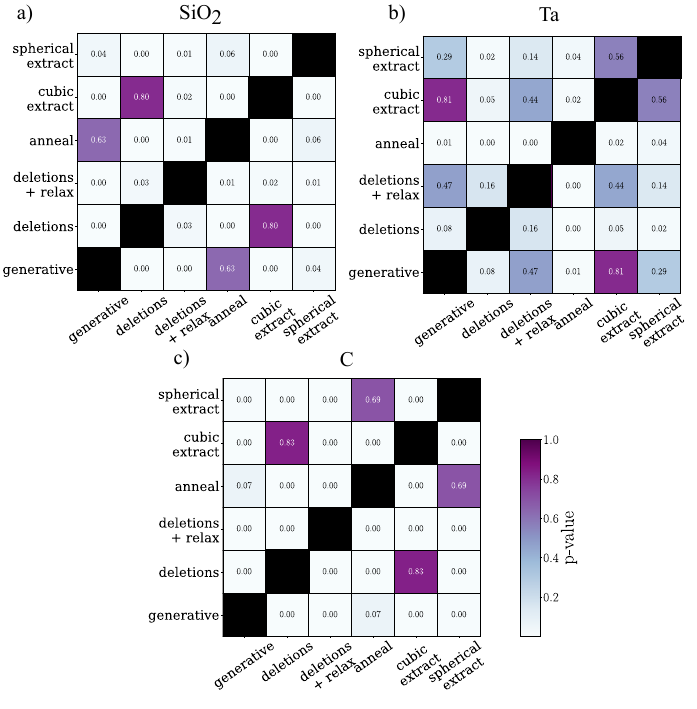}}
\label{fig:stat_sig}
\caption{Paired Student t-test between sets of squared-residual errors, exclusively of the forces on the atoms at the center of the extracted cells}
\end{figure}

\end{document}


\title{Supporting Information for Extracting Atomic Environments for Machine Learning Interatomic Potentials} 

\author{Jared C. Stimac}
\email[]{stimac1@llnl.gov}

\affiliation{Lawrence Livermore National Laboratory \\
  Livermore, California 94550, USA \\}

\author{Fei Zhou}
\affiliation{Lawrence Livermore National Laboratory \\
  Livermore, California 94550, USA \\}

\author{Kyle Bushick}
\affiliation{Lawrence Livermore National Laboratory \\
  Livermore, California 94550, USA \\}

\author{Bo Lei}
\affiliation{Lawrence Livermore National Laboratory \\
  Livermore, California 94550, USA \\}
  
\author{Sebastien Hamel}
\affiliation{Lawrence Livermore National Laboratory \\
  Livermore, California 94550, USA \\}

\author{Amit Samanta}
\affiliation{Lawrence Livermore National Laboratory \\
  Livermore, California 94550, USA \\}

\author{Vincenzo Lordi}
\affiliation{Lawrence Livermore National Laboratory \\
  Livermore, California 94550, USA \\}

\pacs{}

\maketitle 

\section{Repulsive wall potential} 
\label{sec:replusive_wall}
The \textit{deletions + relax} and the \textit{anneal} extraction methods required a repulsive wall potential in order to deter new atoms from entering the spherical, fixed-core region.
It was only applied to atoms that did not start in the fixed-core.
The functional form chosen for this is: 

\begin{equation}
    e_i = 
    \begin{cases}
   \epsilon \left(\dfrac{d^\mathrm{w}_i}{\sigma}\right)^{\mathrm{2}}  & \text{ if} \ d^\mathrm{w}_i > 0, \\
    0                                   & \text{ if} \ d^\mathrm{w}_i \leq 0
    \end{cases}
\end{equation}

where $e_i$ is the energy of atom $i$;  $d^\mathrm{w}_i$ is the $i$-th atom's displacement from the closest point on the wall boundary, with the convention that positions inside the sphere are positive and outside are negative; $\sigma $ is a distance scale, set to 0.5 $ \mathrm{\AA}$; $\epsilon$ is a energy parameter, fixed to $1 \ \mathrm{eV}$.
The boundary of the wall was placed at the boundary of the fixed core for each extracted environment.

\section{Computational cost}
\begin{figure}[!htb]%
\includegraphics[width=0.45\textwidth]{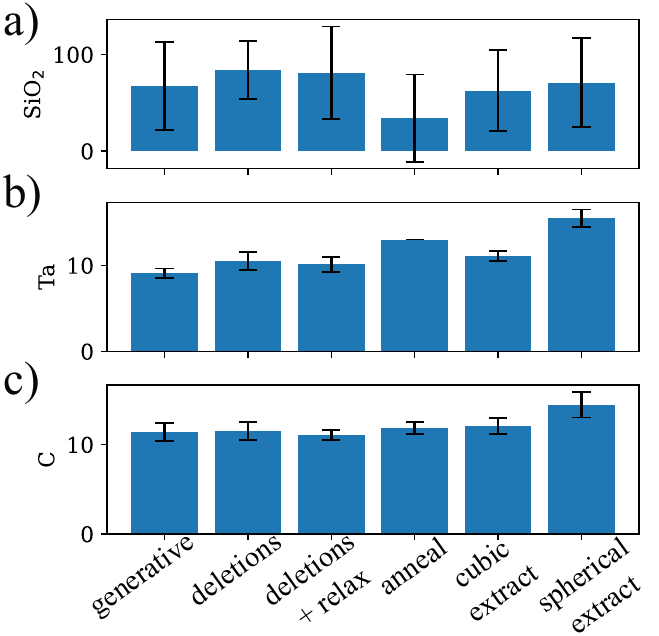}
\caption{Average number of Self-consistent field (SCF) steps for each extraction method and material. Error bars reflect 95\% confidence intervals. }
\label{fig:scf}
\end{figure}

In Fig. S\ref{fig:scf}, the average number of self-consistent field (SCF) steps needed to achieve convergence is reported for each extraction method and material. 
Because the computational resources varied between DFT calculations, we use this quantity as a fair basis for comparing computational cost. 
That said, this only considers if the expense of the DFT single point calculation as a function of the configurations constructed by the extraction methods, and not any computational effort in constructing the configurations. 
For a given material system, the differences in number of SCF steps between the six methods are mostly insignificant.
Two exceptions to this being, first, that \textit{spherical extract} required $\sim30\%$ more steps on average than the other methods in the case of C and Ta, and, second, \textit{Anneal} had far fewer steps than other methods for $\mathrm{SiO_2}$.

\clearpage

\section{Atoms selected for extraction for Ta}
\begin{figure}[!htb]%
\includegraphics[width=0.45\textwidth]{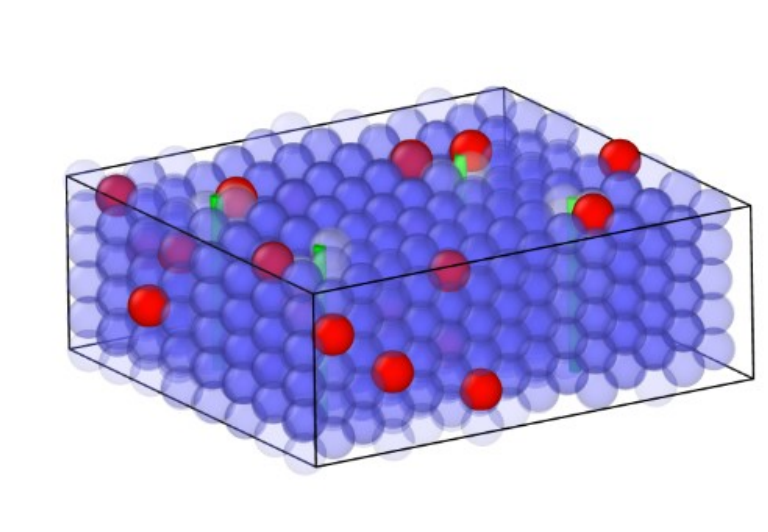}
\caption{Ta source configuration with DXA-evaluated dislocation lines displayed. Red atoms are the 25 selected central atoms for extraction.}
\label{}
\end{figure}

 \newpage

\section{Densities}

\begin{table}[h]
\label{table:densities}
\caption{Average densities in $g/cm^3$ for all configurations for each extraction method as well as that of the source configuration.}
\begin{tabular}{lccc}
\toprule%
                    &       $\mathrm{SiO_2}$ \ \ \   &     Ta   \ \ \  &        C  \ \ \    \\ 
\hline
generative          &           1.86               &   16.30      &       2.32    \\
 deletions          &           2.08                &    15.28     &        2.43   \\
deletions + relax   &           2.08                &    15.28     &         2.43   \\
anneal               &          1.74               &   11.22      &         2.01  \\
cubic extract       &           2.22                &    18.98     &         2.51   \\
spherical extract   &           0.40                &    6.42      &         0.83  \\
\hline
source   (ref.)     &           2.20                   &   16.74       &        2.53
\end{tabular}
\end{table}

\clearpage
\newpage

\section{Zoomed-In View of $\mathrm{SiO_2}$ and C RDFs}

\begin{figure}[!htb]%
\includegraphics[width=0.60\textwidth]{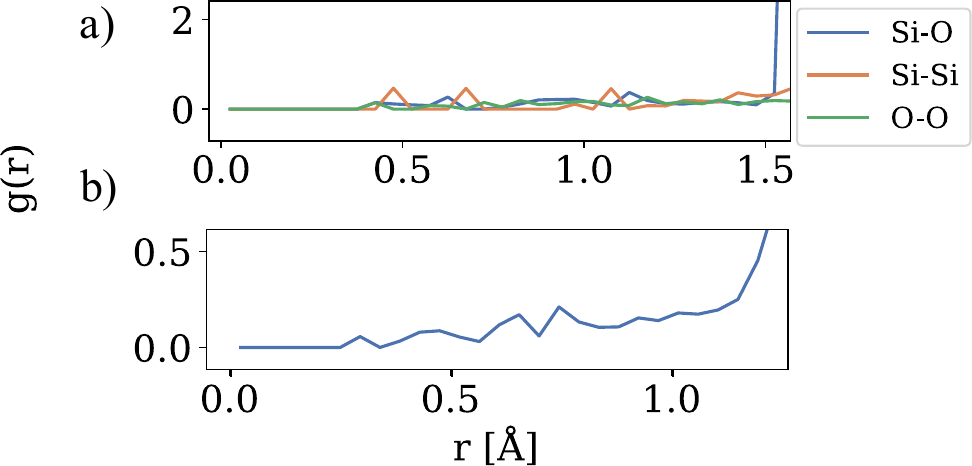}
\caption{RDFs produced by the \textit{cubic extract} method for (a) $\mathrm{SiO_2}$ and (b) C. The magnified view is used to highlight peaks from collisions occurring across the periodic boundaries, which were difficult to see in the full RDFs in the main body of the manuscript.}
\label{fig:stoich}
\end{figure}